\documentclass[review]{elsarticle}
\usepackage{lineno,hyperref}
\usepackage[english]{babel}
\usepackage[numbers]{natbib}
\usepackage{xcolor}
\usepackage{latexsym,amsmath,amssymb,amsbsy,graphicx,geometry}
\modulolinenumbers[5]

\begin{document}
	
	\begin{frontmatter}

\title{Quasi-Stable Structures in Equilibrium Dense Bismuth Melt: Experimental and First Principles Theoretical Studies}

\author[aff1,aff2]{B.N. Galimzyanov}
\ead{bulatgnmail@gmail.com}
\author[aff1]{A.A. Tsygankov}
\author[aff2]{A.V. Suslov}
\author[aff2]{V.I. Lad'yanov}
\author[aff1,aff2]{A.V. Mokshin}

\address[aff1]{Department of Computational Physics, Kazan Federal University, Kazan, Russia}
\address[aff2]{Udmurt Federal Research Center of the Ural Branch of RAS, Izhevsk, Russia}

\begin{abstract}
Near the melting temperature, equilibrium bismuth melt is characterized by structural features that are absent in equilibrium monatomic simple liquids. In the present work, the structure of bismuth melt is studied by X-ray diffraction experiments and quantum chemical calculations. The presence of quasi-stable structures in the melt has been found, the lifetime of which exceeds the structural relaxation time of this melt. It is shown that these structures are characterized by a low degree of ordering and spatial localisation. It was found that up to $50$\% of the atoms in the melt can be involved in the formation of these structures. The elementary structural units of these structures are triplets of regular geometry with the characteristic lengths $3.25$\,\AA~and $4.7$\,\AA~as well as with the characteristic angles $45^{\circ}$ and $90^{\circ}$. The characteristic lengths of these triplets are fully consistent with correlation lengths associated with the short-range order in bismuth melt.
\end{abstract}


\end{frontmatter}

Bismuth is the post-transition metal with application in electronics (printed circuit boards, high and low temperature brazing, etc.), nuclear and solar energy (coolants, targets, concentrators, etc.)~\cite{Kotadia_2014_bi_application1, Osorio_2013_bi_application2, Weeks_1973_bi_nuclear_coolant, Bi_appl3, Bi_appl4, Lorenzin_2016_appl_liquid_metals}. Liquid bismuth as well as its compounds with other metals are used in all these areas. Therefore, the physical and chemical properties of liquid bismuth are crucial factors in the selection of appropriate technologies. The phase diagram of bismuth is quite complex and rich in different crystalline phases: at least five stable crystalline phases are known. At normal pressure, the phase of equilibrium liquid bismuth extends over the temperature range $\Delta T\approx1200$\,K~\cite{Shu_2017}. The transition from crystalline phase to liquid state at normal pressure is accompanied by increasing density~\cite{Plechystyy_2020}.

Near the melting temperature, liquid bismuth has a structure, which differs from the structure typical for the so-called ``simple liquids''. Namely, it demonstrates structural features that manifest itself as a shoulder in the right wing of the main maximum of the static structure factor $S(k)$ as well as an additional peak between the first and second maxima of the radial distribution function $g(r)$. These structural features have been revealed by means of neutron and X-ray diffraction experiments~\cite{Makov_2012,Greenberg_2009_gr_sk}. It is noteworthy that the similar features are also observed in liquid tin and liquid gallium and are explained by the presence of the so-called extended short-range order~\cite{Mokshin_2020_gallium,Mokshin_Novikov_2015}. In Ref.~\cite{Akola_2014}, it was shown that Bi$_{n}$ clusters (where $n=2,\,3,...,14$ is the number of bonded bismuth atoms) can be formed in liquid bismuth. The results of these studies showed that the structural properties (coordination numbers, bond angles, etc.) and energetic properties (including energy barrier and cohesive energy) of these clusters are more similar to those of liquid bismuth than to the stable rhombohedral form of crystalline bismuth. However, there is no convincing evidence that these clusters are long-lived and that these clusters are the cause of the experimentally observed structural features of dense bismuth melt.

The present work will provide evidence that quasi-stable structures are formed in liquid bismuth and that these structures can explain the experimentally observed structural features of liquid bismuth near the melting temperature. For this, detailed study of the structure of liquid bismuth will be carried out using the data from X-ray diffraction experiments and the results of {\it ab-initio} molecular dynamics simulations.

\begin{figure}[ht!]
	\centering
	\includegraphics[width=0.8\linewidth]{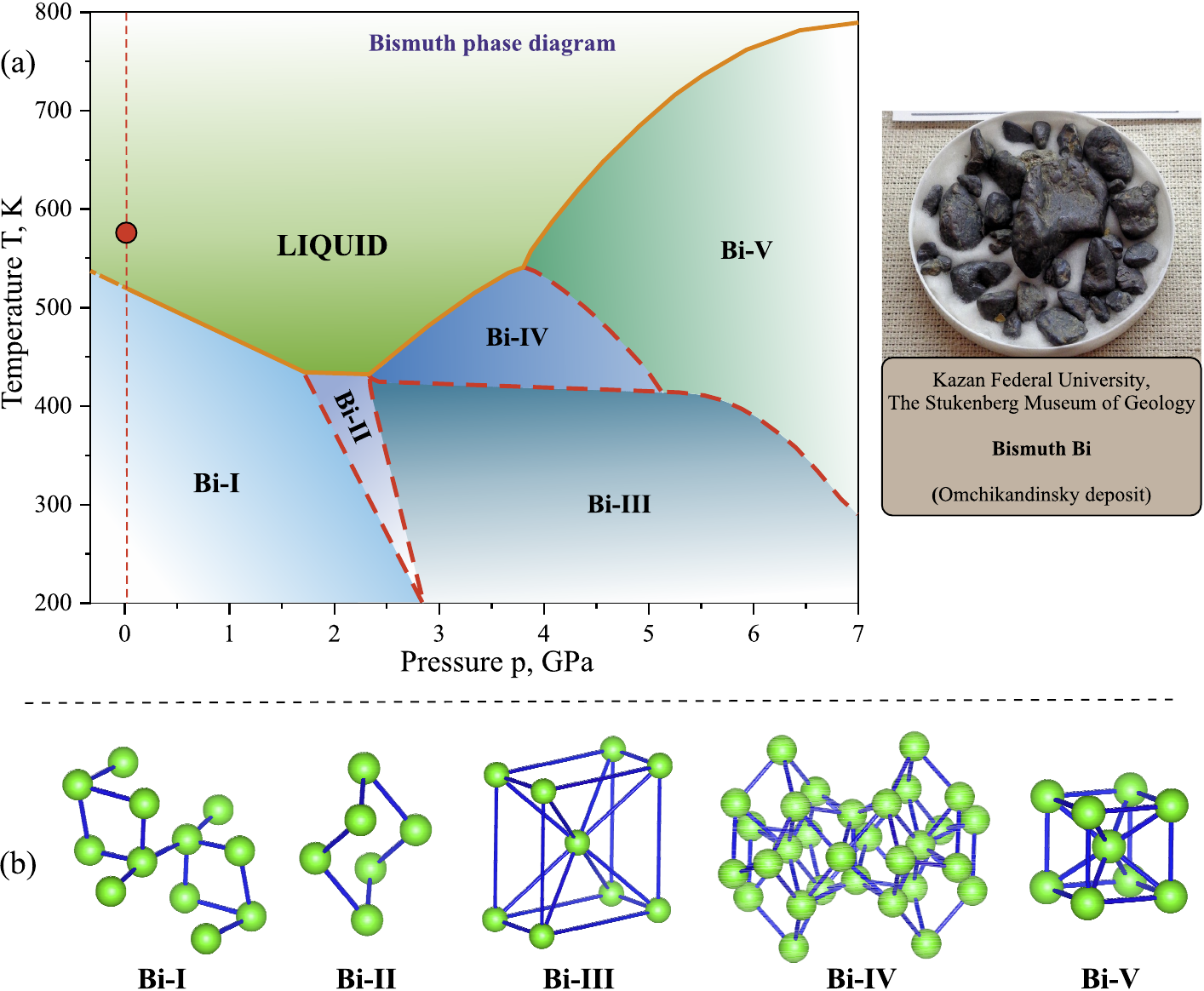}
	\caption{(a) Phase diagram of bismuth created from the data given in Ref.~\cite{Rodriguez_2019_phase_diagram}. The liquid state considered in this study at the temperature $T=573$\,K and at the pressure $p=1$\,atm is marked by the red filled circle. (b) Crystal lattices of different modifications of solid bismuth.}
	\label{bismuth_phase_diagram}
\end{figure}

It is known that the phase diagram of bismuth contains five different crystalline phases in the temperature range $T\in[200; 800]$~K and at pressures up to $7.0$~GPa [see Fig.~\ref{bismuth_phase_diagram}(a)]. A feature of the bismuth phase diagram is that with increasing pressure up to $p\approx2.5$\,GPa the melting point decreases from $T_{m}\simeq545$\,K to $T_{m}\simeq432$\,K. The negative slope of the melting line $T_{m}(p)$ in the ($T$, $p$)-phase diagram reflects the fact that at pressures up to $p\approx2.5$\,GPa the phase transition from crystalline to liquid phase is accompanied by an increase in melt density~\cite{Akola_2014}. The melting line has a positive slope at pressures $p>2.5$\,GPa, where the melting temperature increases with increasing pressure. 

Two rhombohedral crystalline phases with modifications Bi-I and Bi-II are realized at pressures up to $p\simeq2.6$~GPa and at temperatures $T<500$~K [see Fig.~\ref{bismuth_phase_diagram}(b)]. Bi-I phase is stable under normal conditions and this phase is characterized by the following unit cell parameters: $a=b=4.55$~\AA, $c=11.86$~\AA, $\alpha = \beta = 90^\circ$, $\gamma = 120^\circ$~\cite{Akola_2014}. The unit cell of Bi-II phase has the parameters $a = 6.65$~\AA, $b = 6.09$~\AA, $c = 3.29$~\AA, $\alpha = \gamma = 90^\circ$ and $\beta = 110.37^\circ$~\cite{Rodriguez_2019_phase_diagram}. Here, $a$, $b$, $c$ are the edge lengths, $\alpha$, $\beta$, $\gamma$ are the angles between the edges. These two crystalline phases have layered structure, where the bond lengths between atoms in neighboring layers are $\approx3.53$~\AA~(for Bi-I) and $\approx3.45$~\AA~(for Bi-II). Within each layer, some interatomic bond lengths are $\approx3.07$~\AA~(for Bi-I) and $\approx3.94$~\AA~(for Bi-II) that is greater than the doubled covalent radius $2r_{cov}$, where $r_{cov}\simeq1.48$~\AA~\cite{Slater_1964}. Such dissimilarity in the bond lengths between the atoms of crystalline solid bismuth prevents the formation of dense crystal packing. Therefore, Bi-I and Bi-II phases have a lower density than that of the liquid at the common isobar. In addition, crystalline bismuth is characterized by covalent and metallic bonds between the atoms. The results of recent studies show that the covalent bonds are decoupled when crystalline bismuth is melted. Thus, metallic bonds become more pronounced in liquid bismuth~\cite{Kawakita_2018}.

There are high density crystalline phases of solid bismuth with tetragonal, rhombic and cubic lattice at pressures above $p\simeq 2.6$~GPa. In the pressure range from $p\simeq 2.6$ to $7.0$~GPa and at temperatures $T<430$\,K, tetragonal Bi-III phase with the unit cell parameters $a = b = 8.66$~\AA, $c = 4.24$~\AA, $\alpha = \gamma = 90^\circ$ and $\beta = 110.37^\circ$ is formed. High-temperature rhombic Bi-IV phase is realized at pressures in the range $p\in[2.5;\,5. 0]$\,GPa and at temperatures $T\in[430; 550]$\,K, whose unit cell is characterized by the following parameters: $a = 11.19$~\AA, $b = 6.62$~\AA, $c = 6.61$~\AA, $\alpha = \gamma = 90^\circ$, $\beta = 110.37^\circ$. Further, the system forms cubic Bi-V phase with the parameters $a=b=c=3.80$~\AA~and $\alpha = \beta = \gamma = 90^\circ$ at pressures above $p\simeq 4$~GPa and at temperatures above $350$~K. The unit cell images for these five modifications of bismuth crystal lattice are shown in Fig.~\ref{bismuth_phase_diagram}(b).

In the present work, equilibrium bismuth melt is studied just above the melting temperature at the isobar $p=1$~atm ($\sim1.0\times10^{-4}$\,GPa). At this isobar, the liquid phase of bismuth covers the temperature range $T\in[544; 1837]$\,K. The melting temperature is $T_m\simeq544.5$~K, while the boiling temperature is $T_b \simeq 1837$~K~\cite{Lide_2007_book}.
\begin{figure}[ht!]
	\centering
	\includegraphics[width=1.0\linewidth]{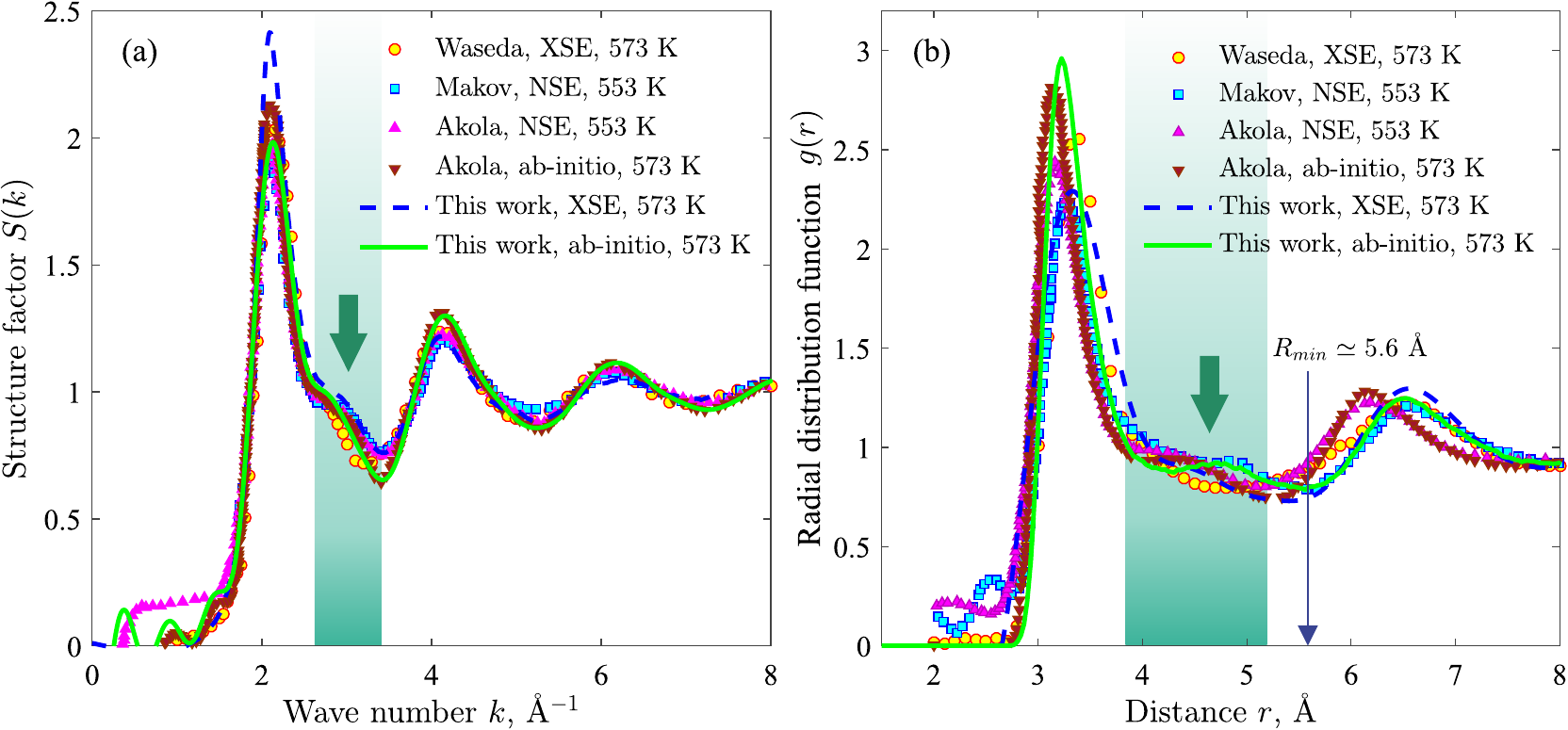}
	\caption{(a) Static structure factor $S(k)$ and (b) radial distribution function $g(r)$ of liquid bismuth at the temperature $T=573$~K obtained from X-ray diffraction experiment and {\it ab-initio} molecular dynamics simulation [see sections ``Experimental procedure'' and ``Simulation procedure'' in Appendix]. Previously known data of X-ray and neutron diffraction experiments as well as known results of quantum chemical calculations are taken from Refs.~\cite{Akola_2014,Greenberg_2009_gr_sk,Makov_2012,Waseda_1972}.}
	\label{gr_and_sk}
\end{figure}

As shown in Figs.~\ref{gr_and_sk}(a) and~\ref{gr_and_sk}(b), the X-ray diffraction results obtained in the present work [see Appendix: ``Experimental procedure''] reveal features in the static structure factor $S(k)$ and the radial distribution function $g(r)$ of liquid bismuth near the melting temperature, which are in agreement with known X-ray (XSE) and neutron (NSE) diffraction data~\cite{Makov_2012,Greenberg_2009_gr_sk,Waseda_1972}. Our results of {\it ab-initio} molecular dynamics simulations are also in agreement with the results of earlier quantum chemical calculations~\cite{Akola_2014}. Fig.~\ref{gr_and_sk}(a) shows that the structural features appear as a shoulder on the right wing of the main peak of $S(k)$. This shoulder is located at the wavenumber interval $k=[2.6;\,3.2]$~\AA$^{-1}$. The radial distribution function $g(r)$ shows an additional peak located between the first and second maxima at values $r=[3.8;~5.2]$~\AA~[see Fig.~\ref{gr_and_sk}(b)].

\begin{figure}[ht!]
	\centering
	\includegraphics[width=1.0\linewidth]{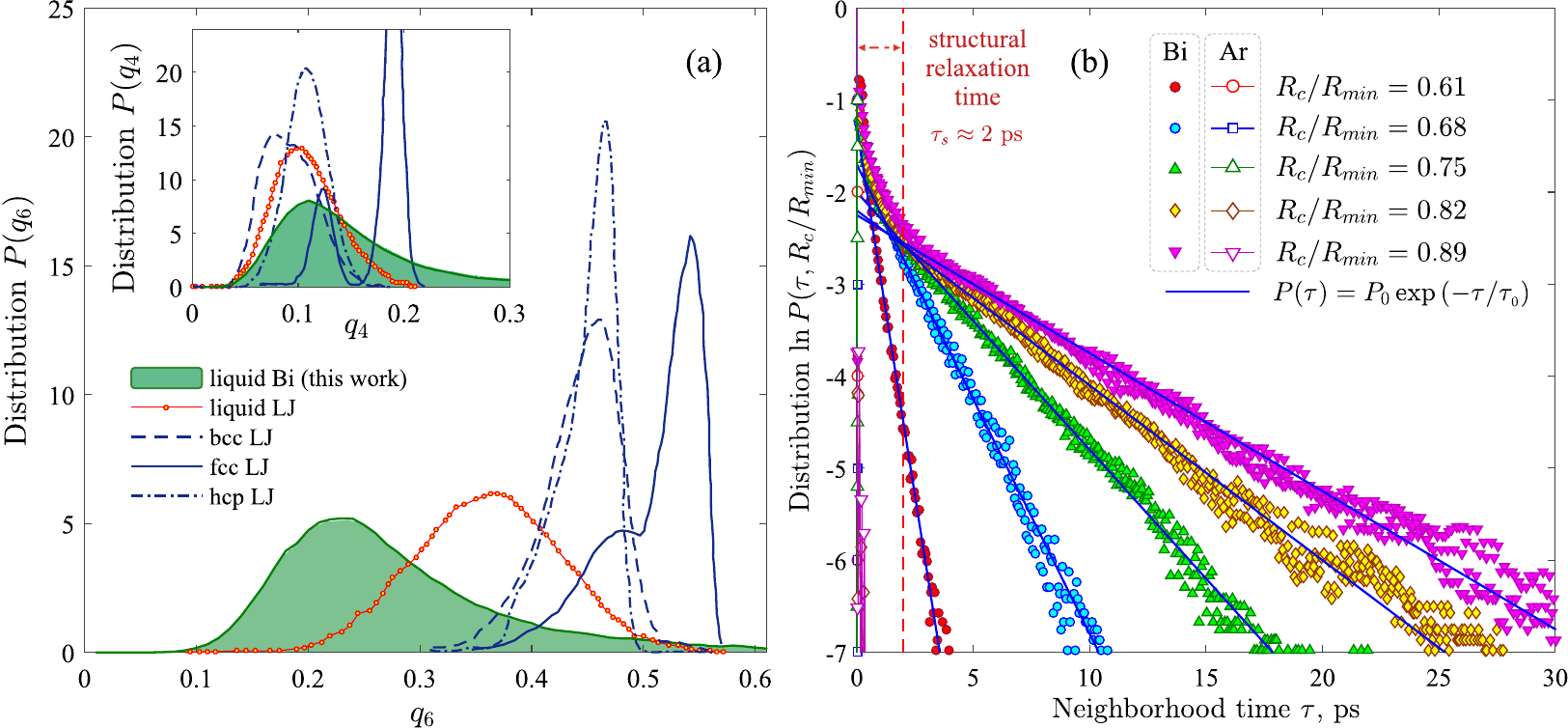}
	\caption{(a) Distributions of the local orientation order parameters $q_4$ and $q_6$ obtained for liquid bismuth as well as previously known results for Lennard-Jones liquid and crystal phases bcc, fcc, hcp~\cite{Lechner_2008}. (b) Distributions of the neighborhood time $\tau$ for pairs of bismuth atoms whose distance is less than the reduced threshold distance $R_c/R_{min}$, where $R_{min}=5.6$\,\AA. For comparison, the figure shows the results obtained for liquid argon at the same values of $T/T_m$ and $R_c/R_{min}$ (here, $R_{min}=5.1$\,\AA~for liquid argon at the reduced temperature $T/T_m=1.05$). The solid lines are the result of the function $P(\tau)=P_{0}\exp(-\tau/\tau_{0})$, where $P_{0}$ is the distribution at the time $\tau=0$; $\tau_{0}$ is the time corresponding to the distribution $P=P_{0}/e\simeq0. 368P_{0}$.} 
	\label{fig_3}
\end{figure}

Fig.~\ref{fig_3}(a) shows the distributions $P(q_4)$ and $P(q_6)$ of the local orientation order parameters $q_4$ and $q_6$ estimated for liquid bismuth at the temperature $T=573$~K  [see section ``Structure and cluster analysis'' in Appendix]. As seen from this figure, the shapes of the distributions $P(q_4)$ and $P(q_6)$ are more asymmetric for liquid bismuth compared to Lennard-Jones liquid~\cite{Lechner_2008}. In the case of liquid bismuth, the position of the maximum in the distribution $P(q_6)$ is shifted to the region of small values $q_6$: the maximum is located at $q_{6}\simeq0.23$, while in the case of Lennard-Jones liquid the maximum is reached at the value $q_{6}\simeq0.37$. Such asymmetric shape of the distributions $P(q_4)$ and $P(q_6)$ as well as the positions of their maxima do not correspond to existing types of the crystal lattice symmetry that can be considered as evidence of the absence of crystal-like clusters in liquid bismuth. The shape of the obtained distributions is completely different from $P(q_4)$ and $P(q_6)$ typical for crystals with bcc, fcc and hcp lattices~\cite{Lechner_2008,Galimzyanov_2019_nucleation}. Namely, in the case of crystalline phases, these distributions have a pronounced maximum or several peaks, the position of which depends on the type of the crystal lattice symmetry~\cite{Mickel_Kapfer_2013}. In the case of liquid bismuth, both distributions $P(q_4)$ and $P(q_6)$ have a unimodal form, typical for ``simple liquids''. 

The distribution $P(\tau,R_c)$ of the neighborhood time $\tau$ for pairs of atoms has been calculated according to the method proposed earlier in Ref.~\cite{Mokshin_2020_gallium}. 
Namely, an imaginary sphere is placed around an arbitrary atom, and the center of the sphere coincides with the center of this atom. The radius of this sphere is defined by the threshold distance $R_c$, which is comparable to the average radius of the atom. This imaginary sphere moves together with the atom according to its dynamics. In addition, the residence times of all atoms that fall into this sphere are determined for the entire observation period. Then, by performing this procedure for all atoms and averaging the results, we obtain the distribution over the neighborhood times of atoms for a given threshold distance $R_c$. As follows from the function $g(r)$ of liquid bismuth [see Fig.~\ref{gr_and_sk}(b)], the structural features appear on the spatial scales [$3.8$; $5.2$]~\AA. With this in mind, the calculation of the neighborhood times of atoms was performed at five different threshold values $R_c=3.4$, $3.8$, $4.2$, $4.6$, and $5.0$~\AA. Here, the threshold distance $R_c=3.4$~\AA~is considered to compare the obtained results with the case, where the structural features are not observed. In Fig.~\ref{fig_3}(b), the obtained distributions are represented as $\ln P(\tau,R_c/R_{min})$ versus $\tau$ plot, where the quantity $R_{min}\simeq5.6$\,\AA~is the position of the first minimum in the function $g(r)$ [see Fig.~\ref{gr_and_sk}(b)]. These distributions have a similar shape and are characterized by two regions: the nonlinear region at small neighborhood times and the linear region associated with the decaying distribution $P(\tau,R_c)$ at large neighborhood times according to the exponential law $P(\tau)=P_{0}\exp(-\tau/\tau_{0})$. Here, the parameters $P_{0}$ and $\tau_{0}$ depend on the threshold value $R_{c}$ [see caption of Fig.~\ref{fig_3}(b)]. At the threshold distances $R_c\geq3.8$~\AA~the transition time between these two regions coincides with the structural relaxation time $\tau_s$ of liquid bismuth at the temperature $T=573$~K. The approximate value of the structural relaxation time $\tau_s$ is determined using the well-known Williams-Landel-Ferry relationship, $\tau_{s}=\tau_{\infty}(\eta/\eta_{\infty})$~\cite{Williams_Landel_1955}. Here, the value of the dynamic viscosity at the temperature $T=573$~K is taken as $\eta\simeq2.18\times10^{-3}$~Pa$\cdot$s~\cite{Greenberg_2009_gr_sk,Chusov_Pronyayev_2020}. The quantities $\eta_{\infty}$ and $\tau_{\infty}$ take the values $\eta_{\infty}\simeq1\times10^{-5}$~Pa$\cdot$s and $\tau_{\infty}\approx0.01$~ps in the case of liquid bismuth at temperatures $T\rightarrow\infty$~\cite{Demmel_Hennet_2021,Hecksher_Dyre_2017}. Then, according to the Williams-Landel-Ferry relationship, the structural relaxation time of liquid bismuth is $\tau_s\simeq2$\,ps at the considered thermodynamic conditions. If pairs of atoms form a stable bond, then the lifetime of this bond is expected to be much longer than the found structural relaxation time of the system. The presence pairs of atoms whose neighborhood time is much longer than $\tau_s\simeq2$\,ps is confirmed by the presence of the linear region in the $\ln P(\tau,R_c/R_{min})$ versus $\tau$ plot. Thus, the quasi-stable structures in liquid bismuth are such structural formations, where interatomic bonds are realized at distances not greater than $5.0$~\AA~and exist longer than $2$\,ps.

For comparison, the distribution of the neighborhood times of atoms was calculated for liquid argon, which belongs to the class of simple monatomic liquids~\cite{Alisson_1967_argon,Wu_1999_MD}. The conditions and parameters were chosen so that the temperature ratio $T/T_m$ and threshold distance $R_c/R_{min}$ for liquid argon coincide with the values of $T/T_m$ and $R_c/R_{min}$ for liquid bismuth. Here, $R_{min}$ is the position of the first minimum of the function $g(r)$: $R_{min}=5.6$\,\AA~for liquid bismuth and $R_{min}=5.1$\,\AA~for liquid argon. Thus, in the case of a simple liquid, quasi-stable structures are not found. This is confirmed by the small neighborhood times of atoms not exceeding $1.0$~ps: in Fig.~\ref{fig_3}(b), the obtained distribution for the simple liquid is located in a narrow region immediately near zero time value. In addition, the distribution for the simple liquid does not have an inflection with transition to the linear regime as it is observed for liquid bismuth at threshold radii $R_c/R_{min}\geq0.68$ [see Fig.~\ref{fig_3}(b)]. 
\begin{figure}[t!]
	\centering
	\includegraphics[width=1.0\linewidth]{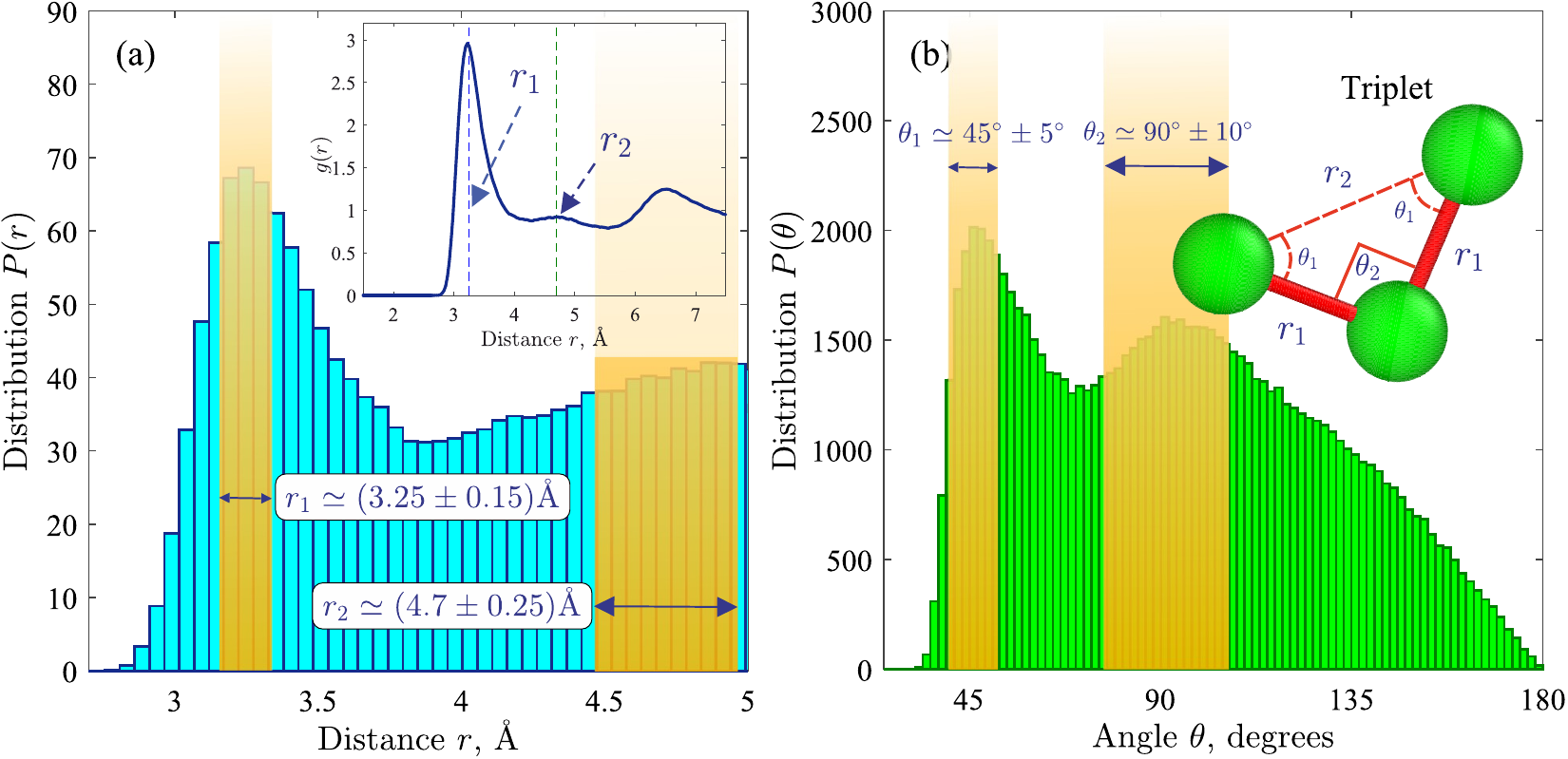}
	\caption{(a) Distribution of the characteristic bond lengths for atoms whose distance does not exceed $R_{c}=5.0$~\AA. Calculations were performed for bonds that live longer than $2$\,ps. Inset shows the radial distribution function $g(r)$ with the marked characteristic lengths $r_{1}$ and $r_{2}$. (b) Distribution of the bond angles between the triplet atoms with the revealed characteristic angles $\theta_1$ and $\theta_2$. Inset shows a schematic view of the triplet with the characteristic lengths $r_{1}$, $r_{2}$ and angles $\theta_1$, $\theta_2$.}
	\label{fig_4}
\end{figure}

A quantitative characterization of the quasi-stable structures is performed by calculation of the distributions of the bond lengths $P(r)$ for pairs of atoms and the bond angles $P(\theta)$ for triples of neighboring atoms belonging to the detected structures. It can be seen from Fig.~\ref{fig_4}(a) that the most probable bonds have the lengths $r_{1}\simeq(3.25\pm0.15)$~\AA~ and $r_{2}\simeq(4.7\pm0.25)$~\AA. In the radial distribution function $g(r)$, these lengths coincide with the position of the main maximum (in the case of $r_{1}$) and the position of the additional peak (in the case of $r_{2}$) [see inset in Fig.~\ref{fig_4}(a)]. The length $r_{1}$ is greater than the doubled covalent radius, $r_{1}>2r_{cov}$, and this length is comparable to the doubled atomic radius, $r_{1}\approx2r_{rad}$, where $r_{rad}=1.6$~\AA~\cite{Slater_1964}. This indicates that the atoms in the quasi-stable structures are bound predominantly by metallic-type bonds. This finding is also in agreement with the results of the recent quasi-elastic neutron scattering measurements, which revealed the absence of covalently bound structures in liquid bismuth~\cite{Kawakita_2018}. These studies are based on the analysis of the characteristic interatomic bond lengths by using the intermediate scattering functions and the van Hove correlation functions.

From the obtained distribution of the bond angles $P(\theta)$ for the triplet atoms [see Fig.~\ref{fig_4}(b)] it follows that the most probable angles are $\theta_1\simeq45^{\circ}\pm4^{\circ}$ and $\theta_2\simeq90^{\circ}\pm10^{\circ}$. The presence of the angle $\theta_2\simeq90^{\circ}$ can be due to electron properties in the outer $6p$ orbital of bismuth atoms, where the valence angle is also $\sim90^\circ$. This valence angle is typical for chemical elements belonging to the pnictogens~\cite{Reimers_2015_bi_bond_angle, Hitomi_2014_bi_bond_angle_book}. Moreover, the characteristic angles $\theta_1$ and $\theta_2$ are typical for all modifications of crystal bismuth including Bi-I phase, which is closest to the considered region of the phase diagram on the isobar $p=1$\,atm~\cite{Rodriguez_2019_phase_diagram}. It follows that quasi-stable structures are formed from triplets acting as elementary structural units. Each triplet is an isosceles triangle with the characteristic angles $45^{\circ}$ and $90^{\circ}$ as well as with the edge lengths $r_{1}$ and $r_{2}$; atoms are placed in the vertices of the triangle [see inset in Fig.~\ref{fig_4}(b)]. The key condition is that the neighborhood time of the atoms in the triplet must exceed $2$~ps. The presence of the triplets is directly manifested in the characteristic shoulder of the function $S(k)$ and in the additional peak of the function $g(r)$ observed from diffraction experiments.

\begin{figure}[h!]
	\centering
	\includegraphics[width=1.0\linewidth]{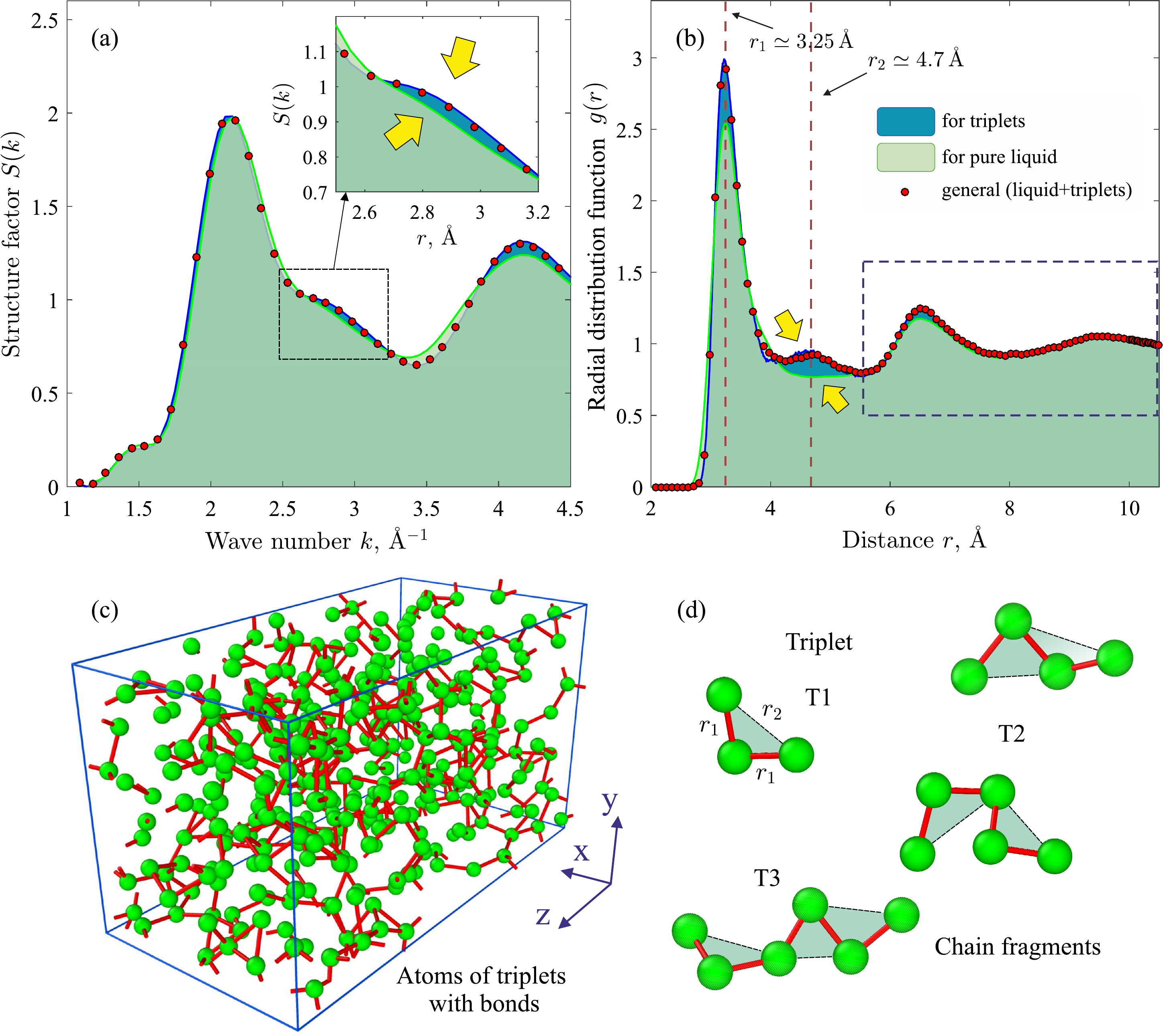}
	\caption{(a) Structure factor $S(k)$ and (b) radial distribution function $g(r)$ calculated separately for the liquid atoms and for the triplet atoms. The circular red markers indicate the total function $g(r)$. (c) Snapshot of the system, where triplet atoms form the branched structure. The elementary unit of this structure is a triplet. (d) Examples of chain fragments formed by bismuth atoms.}
	\label{fig_5}
\end{figure}

For a more detailed interpretation of the shape of the static structure factor and the radial distribution function, the functions $S(k)$ and $g(r)$ are calculated separately for the atoms in the liquid phase without triplets and separately for the atoms forming triplets. As can be seen from Fig.~\ref{fig_5}(a), the shoulder in $S(k)$ is almost absent in the case, where the atoms forming triplets are excluded from consideration. The shoulder is clearly visible in the case of the system consisting only of the triplet atoms; here, the obtained static structure factor approximates the experimental values. Similarly, Fig.~\ref{fig_5}(b) shows that the additional peak in the radial distribution function $g(r)$ is almost absent in the case of the system without triplets. Thus, we conclude that the characteristic shoulder in $S(k)$ and the additional peak in $g(r)$ of liquid bismuth are due to the presence of the quasi-stable structures formed by triplets. The results of the cluster analysis reveal that these triplets are able to form branched structures in the form of chains [see Figs.~\ref{fig_5}(c) and~\ref{fig_5}(d)]. The configuration of these chains can be changed due to thermal motion of the atoms and rearrangement of the triplets. At the same time, the characteristic lengths and angles between the bound atoms are saved, which creates the effect of the presence of quasi-stable structures. At the considered thermodynamic conditions, the fraction of atoms forming these chains can be up to half of the atoms in the system. 

As it is known, the medium-range order means the presence of structures in a liquid whose liner sizes exceed a size of the first coordination shell, but these structures do not form the long-range order typical for crystalline solids~\cite{Lan_2021_medium_range_structure}. This order can be manifested through two possible scenarios. First of all, it can be due to stable quasi-molecular formations, for example, as in the case of supercooled polymer melts. On the other hand, anisotropy in the interparticle interaction can lead to such particle dynamics that creates illusion of structural regularity extending beyond the first coordination. In the case of liquid bismuth near the melting temperature, the detected triplets, as small structural elements, do not form any long-lived structures and, therefore, do not produce the medium-range order in accordance with the first scenario. On the other hand, the medium-range order is also does not manifest itself in such the statistically averaged characteristic as the radial distribution function $g(r)$ of particles. Namely there are no pronounced features in either the second or third peaks of $g(r)$, the specific shape of which could signal the medium-range order [see Fig.~\ref{fig_5}(b)].

The analysis of the local structure of liquid bismuth and the neighborhood time scales of atoms near the melting temperature is performed on the basis of X-ray diffraction experiments and {\it ab-initio} molecular dynamics simulations. The obtained results reveal the existence of quasi-stable structures formed by triplets. It has been established that the appearance of a shoulder in the static structure factor and an additional peak in the radial distribution function is a consequence of the formation of these triplets. We have shown that these triplets can form branched chains of different lengths. The lifetime of these triplets and chains is longer than the structural relaxation time of liquid bismuth at the considered temperature.

This work was supported by the Russian Science Foundation (project no. 19-12-00022, https://rscf.ru/project/19-12-00022/, accessed on 19 May 2023).

\appendix
\section{Experimental procedure}\label{secA1}

The X-ray diffraction patterns of liquid bismuth at the temperature $T=573$~K were obtained using the $\theta$--$\theta$ diffractometer Bruker D8 Advance with thermal chamber HTK1200N. The measurements were performed in the corundum crucible at the dynamic vacuum $10^{-4}$~mmHg and in the pulse set regime with the discrete step $2\theta$: namely, $0.1$ in the interval from $9^{\circ}$ to $81^{\circ}$ and $0.2$ in the interval from $81^{\circ}$ to $123^{\circ}$. The obtained experimental intensity curves have been smoothed by the Savitzky-Golay method followed by the calculation of the structure factor $S(k)$ and the radial distribution function $g(r)$~\cite{Savitzky_Golay_1964,Davis_Farrow_2013}. The experimental measurements are limited by the wavenumber $k = 7.6$\,\AA$^{-1}$, which can lead to additional false oscillations on $g(r)$ due to the small finite limit $S(k)$ in the Fourier transform. Therefore, according to the Kaplow method~\cite{Kaplow_1965}, the wavenumber range for the structure factor was increased to $15$\,\AA$^{-1}$. At the wavenumber $k \approx3$~\AA$^{-1}$, there is a shoulder in the right wing of the main maximum of the structure factor $S(k)$, which is in agreement with the known literature data~\cite{Makov_2012,Greenberg_2009_gr_sk,Waseda_1972,Emuna_2014,Mayo_2013}. In the case of the function $g(r)$, an additional maximum is observed at $r\approx4.5$\,\AA. The values of the maxima of the functions $S(k)$ and $g(r)$ for liquid bismuth at the temperature $573$~K are given in Tables~\ref{tab_1} and~\ref{tab_2}. 
\begin{table}[tbh]
	\centering
	\caption{Values of the structure factor maxima for liquid bismuth at the temperature $573$~K.}	
	\begin{tabular}{ccc}
		\hline
		Maximum & $k_{max}$, \AA$^{-1}$ & $S(k_{max})$  \\
		\hline
		first 	& $2.11\pm0.01$ & $2.41\pm0.01$ \\
		second 	& $4.11\pm0.01$ & $1.22\pm0.01$ \\
		\hline
	\end{tabular}\label{tab_1}
\end{table}
\begin{table}[tbh]
	\centering
	\caption{Values of the maxima of the radial distribution function for liquid bismuth at the temperature $573$\,K: $r_{max}$ and $g(r_{max})$ are the positions of the maxima for the distribution function obtained from the experimental diffraction curve; $r_{max}^{*}$ and $g^{*}(r_{max})$ are the positions of the maxima for the $g(r)$ obtained from the structure factor extended by the Kaplow method~\cite{Kaplow_1965}.}	
	\begin{tabular}{ccccc}
		\hline
		Maximum & $r_{max}$, \AA  & $r_{max}^{*}$, \AA & $g(r_{max})$ & $g^{*}(r_{max})$ \\
		\hline
		first	& $3.32\pm0.01$ & $3.30\pm0.01$ & $2.22\pm0.01$ & $2.21\pm0.01$ \\
		second	& $6.55\pm0.01$ & $6.56\pm0.01$ & $1.27\pm0.01$ & $1.26\pm0.01$ \\
		third	& $9.60\pm0.01$ & $9.62\pm0.01$ & $1.06\pm0.01$ & $1.06\pm0.01$ \\
		\hline
	\end{tabular}\label{tab_2}
\end{table}

\section{Simulation procedure}\label{secA2}

The {\it ab-initio} molecular dynamics simulation of liquid bismuth was performed using the interatomic interaction calculated by the Born-Oppenheimer method with the ultrasoft pseudopotential~\cite{Kresse_1993_1,Kresse_1996_1,Kresse_1996_2,Kresse_1993_2}. The Vienna Ab-initio Simulation Package (VASP) was used to compute the trajectories of bismuth atoms. The cutoff energy for VASP calculation is $400$~eV. We considered the thermodynamic state with the temperature $T=573$~K at the pressure $p=1$~atm. The temperature was controlled by the Nose-Hoover thermostat with the relaxation time $120$~fs. The NVT ensemble was considered, where the system has the constant density $\rho_n = 0.0289$~\AA$^{-3}$ corresponding to the experimental value at the temperature $T=573$~K~\cite{Waseda_1972}. The $432$ atoms were located inside the rectangular simulation cell with the side lengths $L_x = 19.05$~\AA, $L_y = 20.79$~\AA~and $L_z=37.75$~\AA. The simulation time step is $\Delta t=0.003$~ps.

\section{Structure and cluster analysis}\label{secA3}

The radial distribution function $g(r)$ was calculated using the configuration data obtained by the molecular dynamics simulation~\cite{Khusnutdinoff_2018_gr_sk}:
\begin{equation}
	g(r) = \frac{1}{4\pi r^2\rho_n}\left\langle\sum_{i=1}^{N}\frac{\Delta n_{i}(r)}{\Delta r} \right\rangle.
\end{equation}
Here, $\Delta n(r)$ is the number of atoms in a spherical layer of the thickness $\Delta r$, $r$ is the distance between two atoms. The structure factor was determined on the basis of the calculated function $g(r)$ by the Fourier transform~\cite{Mokshin_2018_simple_liquids,Mokshin_2019_sc_relax_theory}:
\begin{equation}
	S(k) = 1 + 4\pi\rho_n\int\limits_{0}^{\infty}r^2\left[g(r) - 1\right]\frac{\sin(kr)}{kr}dr,
\end{equation}
where, $k$ is the wavenumber that takes values in the range $[0.1;\,8.0]$\,\AA$^{-1}$.

Identification of ordered structures was done by calculating local orientational order parameters~\cite{Steinhardt_1983_ql}:
\begin{equation}
	q_{l}(i) = \left(\frac{4\pi}{2l + 1}\sum\limits_{m=-l}^{l}\mid q_{lm}(i)\mid^{2} \right)^{1/2},\,\,\,l = \{4,\,6\},
\end{equation}
where
\begin{equation}
	q_{lm}(i)=\frac{1}{N_b(i)} \sum\limits_{j=1}^{N_b(i)}Y_{lm}(\theta_{ij},\,\phi_{ij}).
\end{equation}
Here, $Y_{lm}(\theta_{ij},\,\phi_{ij})$ are the spherical harmonics, $\theta_{ij}$ and $\phi_{ij}$ are the polar and azimuthal angles, $N_b(i)$ is the number of nearest neighbors of the $i$-th particle. In the case of a three-dimensional single-component system, the local structural order has the $4$-fold and/or $6$-fold orientation symmetry~\cite{Mokshin_2020_gallium,Mickel_Kapfer_2013}. Therefore, as applied to bismuth, it is quite sufficient to calculate the local order parameters $q_{4}$ and $q_{6}$, which allows one to reveal all possible types of crystal lattice symmetry. The cutoff radius at calculation of the local order parameters $q_{4}$ and $q_{6}$ is $5.6$\,\AA.

\end{document}